\documentclass{article}
\usepackage{spconf,amsmath,graphicx}
\usepackage{amssymb}
\usepackage{enumitem}
\usepackage{hyperref}
\usepackage{amsmath}
\usepackage{comment}
\usepackage{amsmath,amssymb,epsfig}
\usepackage{bm}
\usepackage{verbatim}
\usepackage{setspace}
\usepackage{subfig}
\usepackage[x11names]{xcolor}
\usepackage[belowskip=2pt,aboveskip=10pt]{caption}

\ninept
\DeclareMathOperator*{\argminA}{argmin}
\newcommand{\bbR}{\ensuremath{\mathbb{R}}}

\newcommand{\bbN}{\ensuremath{\mathbb{N}}}
\newcommand{\bbC}{\ensuremath{\mathbb{C}}}
\newcommand{\cW}{\ensuremath{\mathcal{W}}}

\newcommand{\bg}{\ensuremath{\mathbf{g}}}

\newcommand{\bh}{\ensuremath{\mathbf{h}}}

\newcommand{\bq}{\ensuremath{\mathbf{q}}}
\newcommand{\bs}{\ensuremath{\mathbf{s}}}
\newcommand{\by}{\ensuremath{\mathbf{y}}}

\newcommand{\be}{\ensuremath{\mathbf{e}}}

\newcommand{\balpha}{\ensuremath{\boldsymbol{\alpha}}}
\newcommand{\btheta}{\ensuremath{\boldsymbol{\theta}}}

\newcommand{\btau}{\ensuremath{\boldsymbol{\tau}}}

\newcommand{\bPhi}{\ensuremath{\mathbf{\Phi}}}
\newcommand{\bTheta}{\ensuremath{\mathbf{\Theta}}}
\newcommand{\bW}{\ensuremath{\mathbf{W}}}
\newcommand{\bX}{\ensuremath{\mathbf{X}}}

\newcommand{\bM}{\ensuremath{\mathbf{M}}}
\newcommand{\bP}{\ensuremath{\mathbf{P}}}
\newcommand{\bH}{\ensuremath{\mathbf{H}}}

\newcommand{\bA}{\ensuremath{\mathbf{A}}}

\newcommand{\bSig}{\ensuremath{\mathbf{\Sigma}}}

\newcommand{\bI}{\ensuremath{\mathbf{I}}}
\newcommand{\bJ}{\ensuremath{\mathbf{J}}}

\newcommand{\bT}{\ensuremath{\mathbf{T}}}
\newcommand{\bU}{\ensuremath{\mathbf{U}}}

\newcommand{\bQ}{\ensuremath{\mathbf{Q}}}

\newcommand{\bzero}{\ensuremath{\mathbf{0}}}

\newcommand{\bcA}{\ensuremath{\boldsymbol{\mathcal{A}}}}
\newcommand{\bcU}{\ensuremath{\boldsymbol{\mathcal{U}}}}

\newif\ifproofread

    \title{Time Delay Estimation from Multiband Radio Channel Samples in Nonuniform Noise\footnotemark}

    \name{Tarik Kazaz, Gerard J. M. Janssen and Alle-Jan van der Veen
    \thanks{This research was supported in part by NWO-STW under 
    	contract 13970 (``SuperGPS'').}	}
    \address{Faculty of EEMCS, Delft University of Technology, Delft, The Netherlands}

\begin{document}
\proofreadfalse
\maketitle


\begin{abstract}
	\noindent The multipath radio channel is considered to have a non-bandlimited channel impulse response. Therefore, it is challenging to achieve high resolution time-delay (TD) estimation of multipath components (MPCs) from bandlimited observations of communication signals. It this paper, we consider the problem of multiband channel sampling and TD estimation of MPCs. We assume that the nonideal multibranch receiver is used for multiband sampling, where the noise is nonuniform across the receiver branches. The resulting data model of Hankel matrices formed from acquired samples has multiple shift-invariance structures, and we propose an algorithm for TD estimation using weighted subspace fitting. The subspace fitting is formulated as a separable nonlinear least squares (NLS) problem, and it is solved using a variable projection method. The proposed algorithm supports high resolution TD estimation from an arbitrary number of bands, and it allows for nonuniform noise across the bands. Numerical simulations show that the algorithm almost attains the Cram\'er Rao Lower Bound, and it outperforms previously proposed methods such as multiresolution TOA, MI-MUSIC, and ESPRIT.
\end{abstract}

\begin{keywords}
time-of-arrival, channel estimation, super-resolution, sparse recovery, multiband sampling, cognitive radio
\end{keywords}

\section{Introduction}
    The first step of time-delay (TD) estimation is an estimation of the multipath components of the underlying communication channel. Since the impulse response of multipath radio channels is considered to be not bandlimited, it is challenging to achieve high resolution TD estimation from bandlimited observations of communications signals. Traditional channel modeling is mainly suited for communication system design, where it is more important to estimate the effects of the channel on the signal to perform equalization, rather than estimating the parameters of the underlying multipath channel. Therefore, radio channels are typically modeled as FIR filters where the time resolution of the channel is inversely proportional to the bandwidth of the signal used for channel probing \cite{witrisal2009noncoherent}. 
	
    Therefore, high resolution channel estimation requires modeling assumptions. Modeling the channel impulse response (CIR) as a sparse sequence of Diracs, time-delay estimation becomes a problem of parametric spectral inference from observed bandlimited signals. Under this assumption, theoretically, it is possible to obtain perfect estimates of the channel parameters from an equally finite number of samples taken in the frequency domain \cite{vetterli2002sampling}.
	
    Many algorithms for TD estimation from frequency domain samples have been proposed in the past. These algorithms are usually based on (i) subspace estimation \cite{van1998joint, li2004super, pourkhaatoun2014high, kazaz2018joint}, (ii) finite rate of innovation \cite{vetterli2002sampling, 6596614, maravic2005sampling}, or (iii) compressed sampling \cite{cohen2014channel, mishali2009blind, mishali2011xampling, zhang2009compressed, gedalyahu2010time, gedalyahu2011multichannel} methods. Few of the previous works \cite{maravic2005sampling, mishali2009blind, mishali2011xampling, gedalyahu2011multichannel} discuss issues related to frequency domain sampling. However, these methods are typically complex for the implementation or not robust to noise. 
	
    The resolution of TD estimation is proportional to the frequency aperture of the samples taken in the frequency domain. To improve the resolution of TD estimation, without arriving at unrealistic sampling rates, multiband channel sampling has been proposed in \cite{wagner2012compressed}. In \cite{kazaz2019multiresolution} a practical multibranch receiver for multiband channel sampling has been proposed. Due to hardware impairments of analog electronics components such as low-noise and power amplifiers, the noise level is typically varying across the receiver branches. 
    
	In this paper, we are interested in a generalized algorithm for high resolution TD estimation from an arbitrary number of sampled bands with nonuniform noise. Following the shift-invariance structure of Hankel matrices formed from the acquired samples, we propose an algorithm for TD estimation based on weighted subspace fitting \cite{viberg1991detection, miesprit1992}. We formulate weighted subspace fitting as a separable nonlinear least squares problem and solve it using the variable projection method \cite{golub2003separable}. To initialize the variable projection method, we use the TD estimate obtained via the multiresolution TOA algorithm \cite{kazaz2019multiresolution}. With this initialization, the iteration of the variable projection method converges very fast, typically within three steps for moderate or high signal-to-noise ratios (SNR).
	
	The resulting algorithm is benchmarked through simulations by comparing its performance with the algorithms proposed in \cite{roy1989esprit, swindlehurst2001exploiting, kazaz2019multiresolution} and the Cram\'er Rao Lower Bound (CRLB). The results show that for low SNR, the proposed algorithm provides better performance than previously proposed algorithms, and it almost attains the CRLB.
	
    \begin{figure*}[t]
		\centering
		\subfloat[]{%
			\includegraphics[trim=1 0 0 1,clip, width=7.8cm]{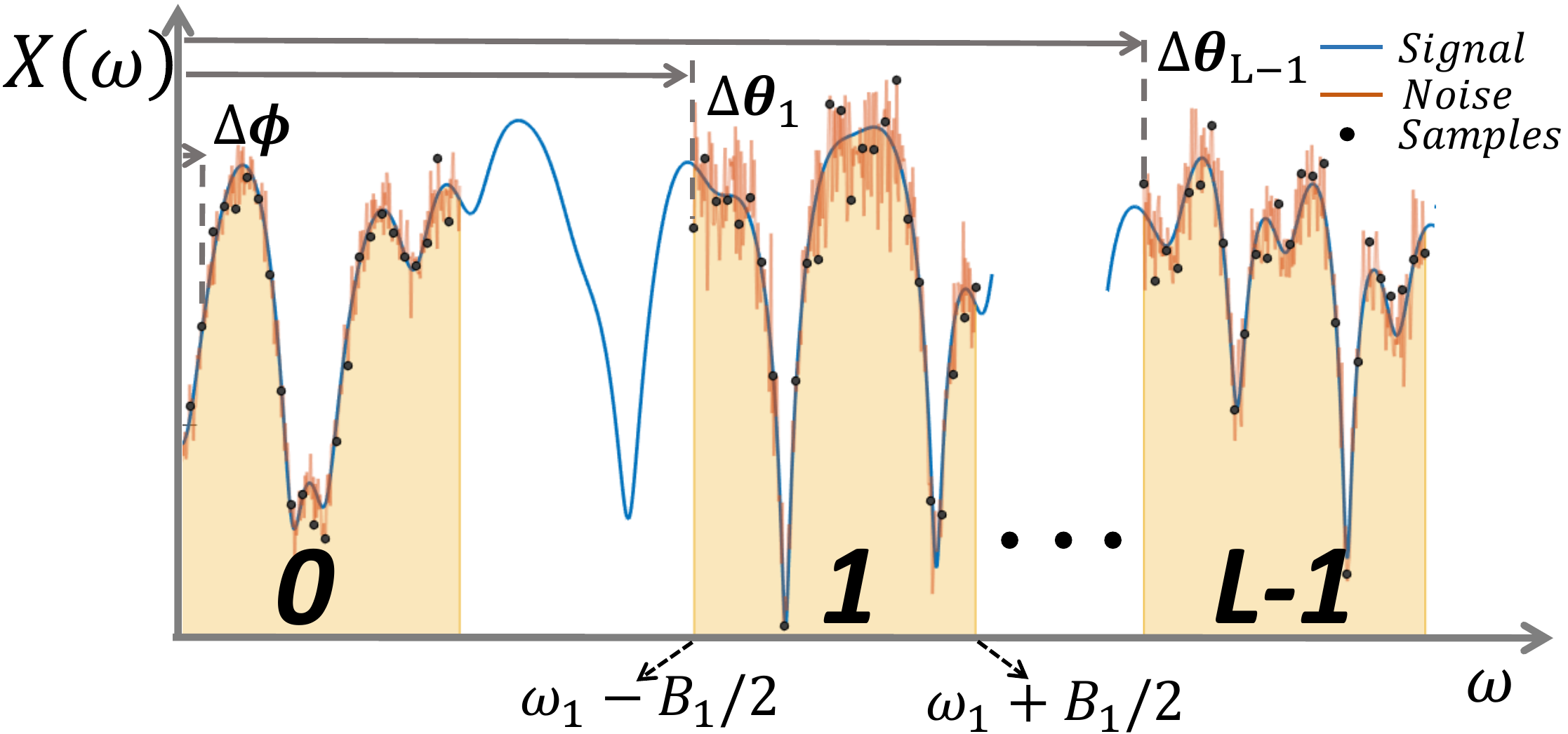}%
			\label{fig:ch:res}%
		}\hspace{2cm} \subfloat[]{%
			\includegraphics[trim=0 1 0 2,clip,width=7.9cm]{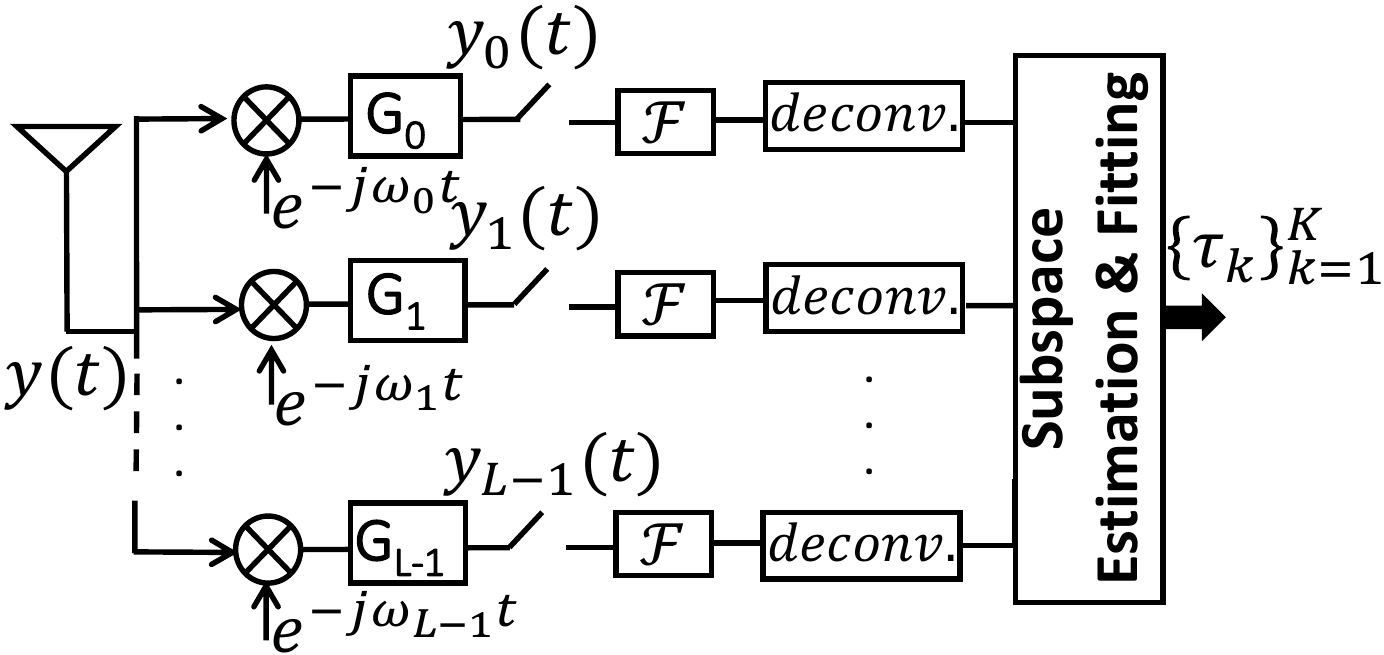}%
			\label{fig:ch:rec}%
		}\qquad
		\caption{(a) The multiband channel frequency response, and (b) a
			multibranch receiver with $L$ RF chains.}
		\label{fig:ch}
    \end{figure*}

	\section{Problem formulation}
    Consider a multipath radio channel model with $K$ propagation paths defined by a continuous-time impulse and frequency response as
    \begin{equation}
    \label{eq:chan_imp}
    h(t) = \sum_{k=1}^{K} \alpha_k \delta(t-\tau_k) 
    \quad \text{and} \quad 
    H(\omega) = \sum_{k=1}^{K} \alpha_k e^{-j\omega\tau_k},
    \end{equation}
    \noindent 
    where $\balpha = [\alpha_1,\dots,\alpha_K]^T \in \bbC^K$ and $\btau = [\tau_1,\dots,\tau_K]^T \in \bbR_+^K$ are collecting unknown gains and time-delays of the MPCs, respectively \cite{molisch2006comprehensive}. We are interested in estimating $\balpha$  and $\btau$ by probing the channel using the known wideband OFDM probing signal $s_i(t)$ transmitted over $i = 0, \dots, L-1$, separate frequency bands (cf. Fig. \ref{fig:ch:res}). The probed frequency bands are $\cW_i = [\omega_i - \frac{B}{2}, \omega_i + \frac{B}{2}]$, where $B$ is the bandwidth, and $\omega_i$ is the central angular frequency of the $i$th band. We consider that measurements are collected using nonideal multibranch transceivers with nonuniform noise across the receiver branches. Our objective is to estimate $\btau$ from an arbitrary number of sampled bands while considering the difference in noise levels of the acquired samples. 
    
    \section{Data model}
    \noindent\textbf{Continous-time signal model:} We consider a baseband signal model and assume ideal conversion to and from the passband without phase and synchronization errors. The response of RF chains at the probed frequency bands are modeled using equivalent linear and time-invariant low-pass filters ${g_i(t)}$, where the corresponding CTFT $G_i(\omega)$ has passband $[-\frac{B}{2}, \frac{B}{2}]$ (cf. Fig. \ref{fig:ch:rec}). We assume that the frequency responses of the RF chains $G_i(\omega)$ are characterized during calibration and known. The algorithm for time-delay estimation in the case when $G_i(\omega)$, $i = 0, \dots, L-1$, are unknown is proposed in \cite{kazaz2019jointcal}.  The impulse response of the $i$th channel band is $h_i(t)$, and its CTFT is $H_i(\omega) = H(\omega_i+\omega)$. Assume that $h_i(t)$ is time-limited to the duration of the OFDM symbol cyclic prefix, that is $h_i(t) = 0$ for $t \notin [0, T_{CP}]$. Therefore, there is no inter-symbol interference, allowing us to consider the model for a single OFDM symbol only.
    
    Consider that the same OFDM probing signal is transmitted in all bands, that is, $s_i(t) = s(t)$ and $B_i = B$ for $i = 0, \dots, L-1$, and $s(t)$ is given by
    \begin{equation}
        \nonumber
        \label{eq:pilot_sig}
           s(t)  =
           \begin{cases}
           \sum_{n=0}^{N-1} s_n e^{j \omega_{sc}nt}, & t \in [-T_{CP}, T_{sym}]
           \\
           \mbox{0}, & \text{otherwise}\,,
           \end{cases}
    \end{equation}
    where $\bs = [s_0,\dots,s_{N-1}]^T \in \bbC^N$ are the known pilot symbols, $\omega_{sc}$ is the sub-carrier spacing, and \(T_{sym} = 2\pi/\omega_{sc}\) is the duration of one OFDM symbol.The signal received at the $i$th band after conversion to the baseband and low-pass filtering is 
    \begin{equation}
        \label{eq:rx_six}
            y_i(t) = s(t) \ast g_i(t) \ast h_i(t) + q_i(t)\,,
    \end{equation}
    where $q_i(t)$ is low-pass filtered Gaussian white noise. The corresponding CTFT of the signal $y_i(t)$ is 
    \begin{equation}
        \label{eq:freq_rx_sig}
           Y_i(\omega)  = 
           \begin{cases}
           S(\omega) G_i(\omega) H_i(\omega)  + Q_i(\omega), & \omega \in [-\frac{B}{2}, \frac{B}{2}]
           \\
           \mbox{0}, & \text{otherwise}\,,
           \end{cases}
    \end{equation}
    \noindent where $S(\omega)$ and $Q_i(\omega)$ are the CTFTs of $s(t)$ and $q_i(t)$, respectively.
    
    \noindent\textbf{Discrete-time signal model:} The receiver samples $y_i(t)$ with period $T_s=1/B$, performs packet detection, symbol synchronization, and removes the cyclic prefix. During the period of single OFDM symbol, $N$ complex samples are collected, where $N$ is the number of sub-carriers and $T_{sym} = NT_s$. Next, $N$-point DFT is applied on the collected samples, and they are stacked in increasing order of the DFT frequencies in $\by_i \in \bbC^{N}$. Then, the discrete data model of the received signal (\ref{eq:freq_rx_sig}) can be written as
    \begin{equation}
        \label{eq:data_model}
            \by_i = \text{diag}(\bs \circ \bg_i) \bh_i + \bq_i\,,
    \end{equation}
    \noindent where $\circ$ is the Khatri-Rao product, $\bs$ collects the known pilot symbols, $\bg_i$, and $\bq_i$, collect samples of $G_i(\omega)$, and $Q_i(\omega)$ at the subcarrier frequencies, respectively. Likewise, $\bh_i \in \bbC^{N}$ collects samples of $H_i(\omega)$ as 
    \begin{equation}
        \label{eq:ch_dis1}
            H_i[n] = H\left(\omega_i + n\omega_{\rm sc} \right), \quad n = -\frac{N}{2}, \dots, \frac{N}{2}\,,
    \end{equation}
    \noindent where $\omega_{\rm sc} = \frac{2\pi}{NT_s}$, and we assume that $N$ is an even number. We consider that bands $\left\{\cW_i\right\}_{i=0}^{L-1}$ are laying on the discrete frequency grid $\omega_i = \omega_0 + n_i\omega_{\rm sc}$, where $n_i \in \bbN$, and $\omega_{0}$ denotes the lowest frequency considered during channel probing. Inserting the channel model (\ref{eq:chan_imp}) into (\ref{eq:ch_dis1}) gives
    \begin{equation}
        \label{eq:ch_dis2}
            H_i[n] = \sum_{k=1}^{K} \alpha_k e^{-jn_i\omega_{sc}\tau_k}e^{-jn\omega_{sc}\tau_k},
    \end{equation}
    \noindent where we absorbed $e^{-j\omega_{0}\tau_k}$ in $\alpha_k\,\forall\,k$. Now, the channel vector $\bh_i$ satisfies the model
    \begin{equation}
    \label{eq:ch_model}
        \bh_i = \bM\text{diag}(\btheta_i)\balpha\,,
    \end{equation} 
    where  $\bM \in \bbC^{N\times K}$ is a Vandermonde matrix
    \begin{equation}
	\label{eq:f_matrix}
	\bM = 
	\begin{bmatrix}
	1 & 1 & \cdots & 1 \\
	\Phi_1 & \Phi_2 & \cdots & \Phi_K\\
	\vdots & \vdots & \ddots & \vdots \\ 
	\Phi_1^{{N-1}} & \Phi_2^{N-1} & \cdots & \Phi_K^{N-1}
	\end{bmatrix}
	\,,\quad
	\end{equation}
	and $ \Phi_k = e^{-j\phi_k}$, where $\phi_k = \omega_{sc} \tau_k$. Likewise, the band dependent phase shifts of MPCs are collected in $\btheta_i = [\theta_{i,1},\dots,\theta_{i,K}]^T \in \bbC^K$, where $\theta_{i,k} = \Phi_k^{n_i}$.
	
	Next, we assume that none of the entries of $\bs$ or $\bg_i$ are zero or close to zero, and estimate $\bh_i$ by applying deconvolution on the data vector (\ref{eq:data_model}) as 
	\begin{equation}
    \nonumber
        \label{eq:ch_est}
            \bh_i = \text{diag}^{-1}(\bs\circ\bg_i)\by_i\,,
    \end{equation}
    \noindent which satisfies the model 
    \begin{equation}
        \label{eq:ch_est_model}
            \bh_i = \bM\bTheta_i\balpha + \bq_i'\,.
    \end{equation}
    \noindent where $\bTheta_i = \text{diag}(\btheta_i)$. The pilot symbols $\bs$ have the constant magnitude, and we assume that frequency responses of receiver chains $\bg_i$ are almost flat. Therefore $\bq_i'= \text{diag}^{-1}(\bs\circ\bg_i)\bq_i$ is a zero-mean white Gaussian distributed noise with covariance $\bSig_{i} = \sigma_{i}^2\bI_N$, where $\bI_N$ is the $N \times N$ identity matrix. In the case that the frequency responses of the RF chains are not perfectly flat, $\bq_i'$ will be colored noise. However, its coloring is known and can be taken into account. 

    \section{Multiband time-delay estimation}
	Our next objective is to estimate $\btau$ from the channel estimates $\bh_i$, $i = 0, \dots, L-1$. We begin by stacking the channel estimates in the multiband channel vector $\bh = [\bh_0^T,\dots,\bh_{L-1}^T]^T \in \bbC^{NL}$. From (\ref{eq:ch_est_model}), $\bh$ satisfies the model 
	\begin{equation}
        \label{eq:multiband}
            \bh = \bA(\btau)\balpha + \bq := 
     \begin{bmatrix} 
          \bM \\
          \bM\bTheta_1\\
          \vdots\\
          \bM\bTheta_{L-1}\\
      \end{bmatrix}\balpha + 
      \begin{bmatrix}
          \bq_{0}' \\ \bq_{1}' \\ \vdots \\ \bq_{L-1}'
      \end{bmatrix} \,.
    \end{equation}
    Since $\bA(\btau)$ has a multiple shift-invariance structure and (\ref{eq:multiband}) resembles the data model of Multiple Invariance ESPRIT \cite{miesprit1992}, $\btau$ can be estimated using a subspace fitting methods. 
    
    \subsection{Multiband estimation algorithm}
    From the vectors $\bh_i$, $i=0, \cdots, L-1$, we construct Hankel matrices of size $P\times Q$ as
    \begin{equation}
    \label{eq:hankel}
    \bH_i = \begin{bmatrix}
    H_i[0] & H_i[1] & \cdots & H_i[Q] \\
    H_i[1] & H_i[2] & \cdots &   H_i[Q+1] \\
    \vdots &  \vdots   &   \ddots     & \vdots  \\
    H_i[P-1] & H_i[P] & \cdots & H_i[N-1]
    \end{bmatrix} \,.
    \end{equation}
    Here, $P = N-Q-1$, $Q$ is a design parameter and we require $P>K$ and $Q\ge K$.
    From (\ref{eq:ch_est_model}), and using the shift invariance of the Vandermonde
    matrix (\ref{eq:f_matrix}), the constructed Hankel matrices satisfy
    \begin{equation} 
    \label{eq:hen_mat}
    \bH_i = \bM' \bTheta_i \bX + \bQ_i \,,
    \end{equation} 
    where $\bM'$ is an $P\times K$ submatrix of $\bM$,
    \[
    \bX = [\balpha,\, \bPhi \balpha,\, \bPhi^2 \balpha, \cdots, \bPhi^{Q-1}
    \balpha], \,
    \]
    $\bQ_i$ is a noise matrix, and ${\bPhi = \text{diag}([\Phi_1 \cdots \Phi_K])}$.
    
    The column subspaces of $\bH_i$, $i=0, \cdots, L-1$, are spanned by the same $K$ dimensional basis. Therefore, a good initial estimate of the orthonormal basis that spans the column subspace of $\bH_i$, $i=0, \cdots, L-1$, can be obtained from the low-rank approximation of the block Hankel matrix
    \begin{equation}
    \label{eq:rb_H}  
    \bH_{r} =
    \begin{bmatrix} \bH_0 & \bH_1 & \cdots  & \bH_{L-1} \end{bmatrix}.
    \end{equation} 
    The matrix $\bH_{r}$ satisfies the model
    \begin{equation}
    \label{eq:rb_H_fac} 
    \nonumber 
    \bH_{r} = \bM' \bX_{r} + \bQ_{r},
    \end{equation}
    where $\bX_{r} = [\bX,\, \bTheta_1\bX ,\, \cdots,\, \bTheta_{L-1}\bX]$, and likewise $\bQ_{r} = [\bQ_0,\, \cdots,\, \bQ_{L-1}]$. Let $\bU_r$ be a $K$ dimensional orthonormal basis for the column span of $\bH_{r}$, then $\bP_{\bU_r} = \bU_r\bU_{r}^H$ is the corresponding projection matrix. To perform noise reduction, we project $\bH_i$, $i=0, \cdots, L-1$, onto the column subspace of $\bU_r$ and form the block Hankel matrix 
    \begin{equation}
    \label{eq:cb_H} 
    \nonumber 
    \bH = \left(\bI_L \otimes \bP_{\bU_r}\right)
    \begin{bmatrix} \bH_0 \\ \bH_1 \\ \vdots\\ \bH_{L-1} \end{bmatrix}\,,
    \end{equation} 
    where $\otimes$ is the Kronecker product. Now, $\bH$ satisfies the model
    \begin{equation} 
    \label{eq:cb_H_fac} 
    \bH = \bA'(\btau) \bX  + \bQ := 
        \begin{bmatrix} 
          \bM' \\
          \bM' \bTheta_1\\
          \vdots\\
          \bM' \bTheta_{L-1}\\
       \end{bmatrix} \bX +
       \begin{bmatrix} 
          \bP_{\bU_r} \bQ_0 \\
          \bP_{\bU_r} \bQ_1\\
          \vdots\\
          \bP_{\bU_r} \bQ_{L-1}\\
        \end{bmatrix}\,.
    \end{equation}
    
    Note that $\bA'(\btau)$ has multiple shift-invariance structures introduced by the phase shifts of $\btau$ on the (i) subcarrier frequencies of the pilots, $\bPhi$, and (ii) carrier frequencies of the bands, $\bTheta_i = \bPhi^{n_i}$, $i=0, \cdots, L-1$, as shown in Fig.\ref{fig:ch:res}. The phase shifts $\bPhi$ can be estimated from the low-rank approximation of $\bH$ and its shift-invariance properties using subspace fitting methods. From the estimate of $\bPhi$, $\btau$ immediately follows. 
    
    Let $\bU$ be a $K$-dimensional orthonormal basis for
    the column span of $\bH$, obtained using the singular value
    decomposition \cite{van1993subspace}, then we can write $\bA'(\btau) = \bU \bT$, where $\bT$ is a $K\times K$ nonsingular matrix. Next, let us define selection matrices 
    \begin{equation}
	   \label{eq:sel_mat}
	   \begin{aligned} \centering 
	   \bJ_{i,1}  &= \be_{i}^{T} \otimes [\bI_{P-1} \quad \bzero_{P-1} ]\,,\\
	   \bJ_{i,2}  &= \be_{i}^{T} \otimes [\bzero_{P-1} \quad \bI_{P-1}]\,,
	   \end{aligned}
   \end{equation} 
   \noindent where $(\cdot)^T$ is transpose, $\bzero_{P-1}$ is a zero vector of size $P-1$, $\be_i$ is a vector of size $L$, with $i$th element equal to $1$ and zero otherwise. To estimate $\bPhi$ we select submatrices consisting of the first and, respectively, the last row of each block matrix stacked in $\bU$, that is $\bU_{i,1} = \bJ_{i,1} \bU$ and $\bU_{i,2} = \bJ_{i,2} \bU$, $i=0, \cdots, L-1$. In the view of shift-invariance structure of $\bA'(\btau)$, we have
   \begin{equation}
	   \label{eq:sel_sub}
	   \bU_{i,1}  = \bM'' \bPhi^{n_i}\,,\qquad
	   \bU_{i,2}  = \bM'' \bPhi^{n_i+1}\,,
   \end{equation} 
   \noindent where $\bM''$ is a $(P-1)\times K$ submatrix of $\bM'$. Next, we form block matrices \begin{equation} 
    \label{eq:block_matrix} 
    \bcU = 
        \begin{bmatrix} 
          \bU_{0,1} \\
          \bU_{0,2} \\
          \vdots\\
          \bU_{L-1,2}\\
       \end{bmatrix}\,, \qquad
    \bcA = 
        \begin{bmatrix} 
          \bM'' \\
          \bM''\bPhi \\
          \vdots\\
          \bM''\bPhi^{n_{L-1}+1}\\
       \end{bmatrix}\,.
    \end{equation}
    Finally, $\bPhi$ can be estimated by solving the following weighted subspace fitting problem
    \begin{equation} 
    \label{eq:sub_fit_prob}
    \begin{aligned} \centering
    \hat{\bPhi} = & \argminA_{\bPhi} \left\| {\bW}^{1/2} \left(
    \bcU - \bcA \bT^{-1} \right) \right\|_{F}^2 \\
    = &\argminA_{\bPhi} \left\| {\bW}^{1/2} 
    \left(\bI-\bP_{\bcA}(\bPhi)\right)\bcU \right\|_{F}^2
    \end{aligned}
    \end{equation}
    
    \begin{figure*}[t!]
		\centering
		\subfloat[]{%
			\includegraphics[trim=0 1 1 2,clip, width=6.8cm]{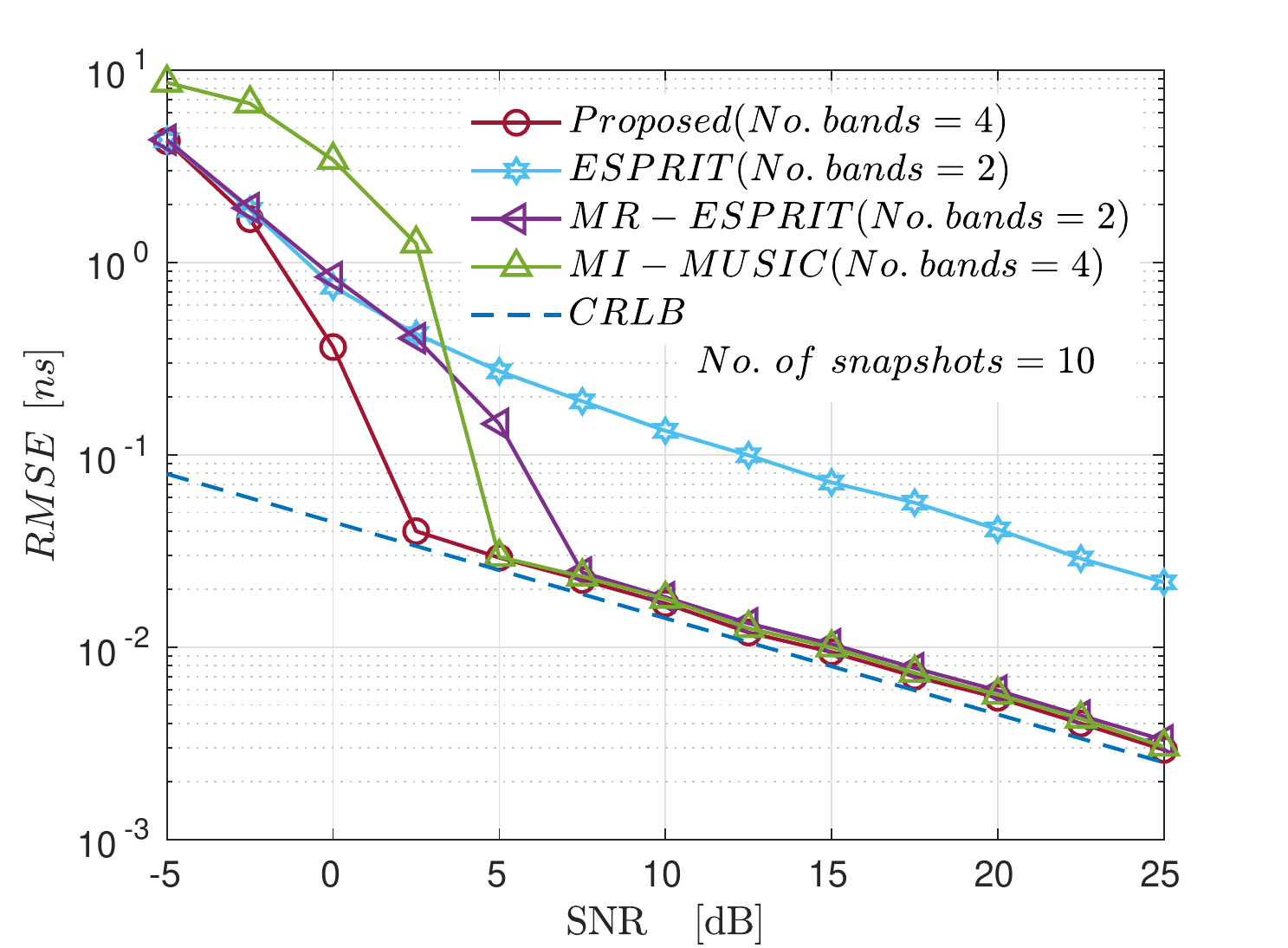}%
			\label{fig:tdu:snr}%
		}\hspace{2cm} \subfloat[]{%
			\includegraphics[trim=0 1 1 2,clip,width=6.8cm]{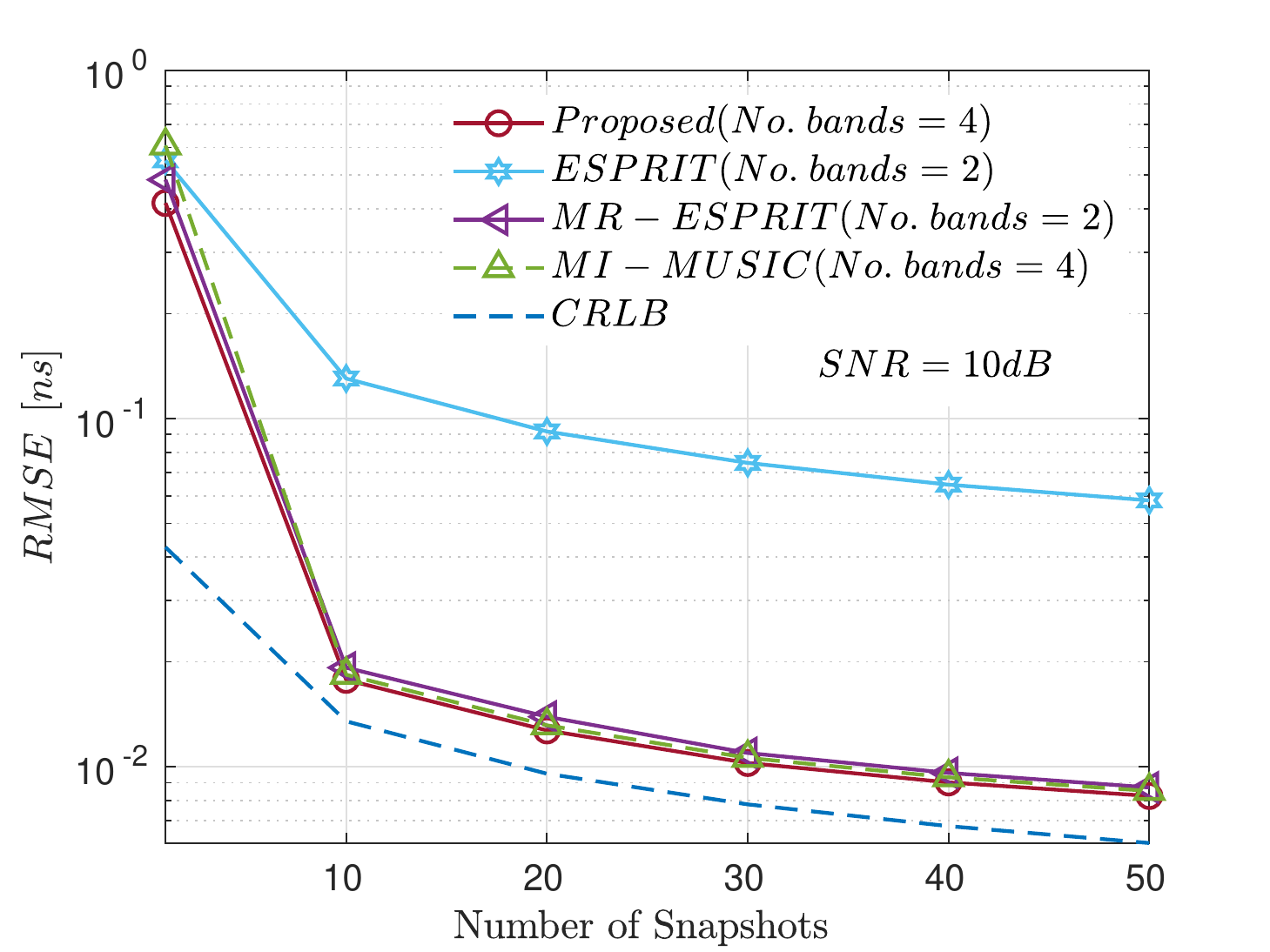}%
			\label{fig:tdu:snap}%
		} \qquad \subfloat[]{%
			\includegraphics[trim=0 1 1 2,clip,width=6.8cm]{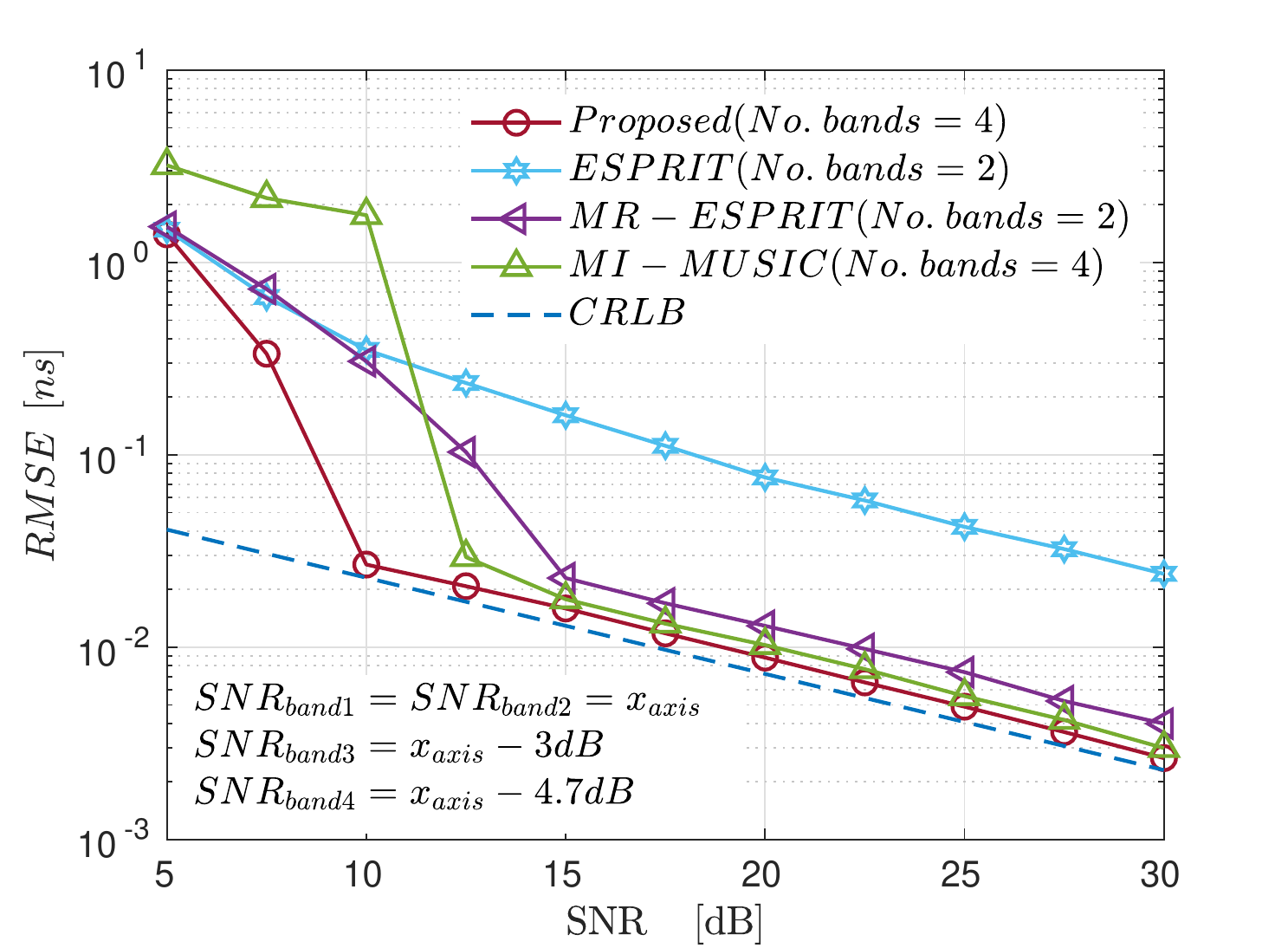}%
			\label{fig:tdnu:snr}%
		} \hspace{2cm} \subfloat[]{%
			\includegraphics[trim=0 1 1 2,clip,width=6.8cm]{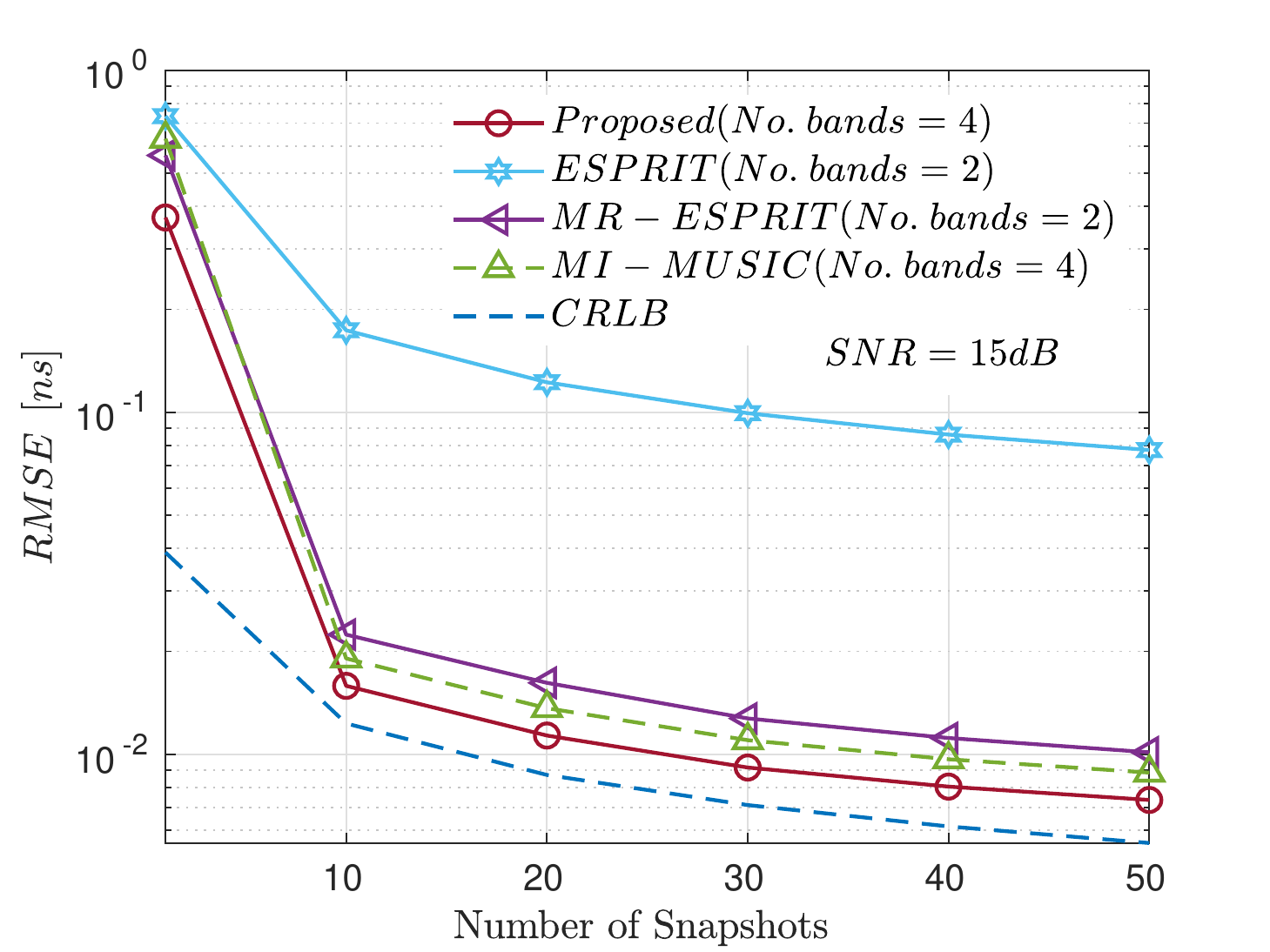}%
			\label{fig:tdnu:snap}%
		} \qquad
		\caption{RMSE for estimated time-delay in uniform (a-b) and nonuniform noise (c-d) vs signal to noise ratio and number of snapshots.} 
		\label{fig:ch}
    \end{figure*}
    \noindent where $\bP_{\bcA}(\bPhi) = \bcA\bcA^{\dagger}$, $(\cdot)^{\dagger}$ is the pseudoinverse of a matrix, and $\bW = \bI_{2L(P-1)}$ for the case when noise is uniform accross the receiver chains branches. When the noise power is nonuniform accross the receiver branches, the weighting martix is given by $\bW =  \text{blkdiag} \left({\bSig'_{1}}^{-1}, \dots , {\bSig'_{L}}^{-1}\right) \in \bbC^{2L(P-1) \times 2L(P-1)}$, where $\bSig'_{i} = \sigma_i^2\bI_{P-1}$ and we assume that $\sigma_i$, $i=0, \cdots, L-1$, are known.

   The subspace fitting problem (\ref{eq:sub_fit_prob}) can be formulated as a separable nonlinear least squares problem, which can be solved efficiently using several iterative optimization methods (e.g., variable projection, Gauss-Newton or Levenberg-Marquardt). We use the variable projection method \cite{golub2003separable} to find a solution, where a good initialization is obtained by the multiresolution TOA algorithm \cite{kazaz2019multiresolution}. With this initialization, the variable projection method converges very fast, typically within three steps for moderate and high signal-to-noise ratios (SNR). 
    
   \section{Numerical Experiments}
   This section evaluates the performance of the proposed algorithm via numerical simulations. We consider a scenario where the multipath channel has nine dominant MPCs, i.e., $K = 9$, with gain of line-of-sight (LOS) MPC distributed according to a Rician distribution.  The continuous-time channel is modeled using a $2$ GHz grid, with channel tap delays spaced at $500$ ps. We consider that the receiver estimates the channel frequency response in four frequency bands, i.e., $L = 4$, using probing signal with $N = 256$ subcarriers and bandwidth of $B = 80$ MHz. The central frequencies of the bands are $\{60, 180, 290, 400 \}$ MHz, respectively. To evaluate the performance of TD estimation, we use the root mean square error (RMSE) of the LOS multipath component TD estimate. The RMSEs are computed using $10^3$ independent Monte-Carlo trials and compared with the CRLB and  RMSEs of the algorithms proposed  in  \cite{kazaz2019multiresolution, roy1989esprit, swindlehurst2001exploiting} which are shortly denoted with MR-ESPRIT, ESPRIT, and MI-MUSIC, respectively. 
   
   Fig.\ \ref{fig:tdu:snr} shows the performance of the proposed, MR-ESPRIT, ESPRIT, and MI-MUSIC algorithms in different signal-to-noise ratio (SNR) regimes for the case when noise power is uniform across the receiver branches. The number of channel snapshots is set to $10$ and kept fixed during trials. From Fig.\ \ref{fig:tdu:snr}, we observe that the RMSE of TD estimation decreases with SNR. The MR-ESPRIT and ESPRIT algorithms utilize only samples available from the first and fourth band. Therefore, as expected, their performance is worse, compared to the performance of the proposed and MI-MUSIC algorithms. The MR-ESPRIT, MI-MUSIC and the proposed algorithms are all almost attaining the CRLB for sufficiently high SNR, while ESPRIT due to inefficient use of available data is not able to resolve closely spaced MPCs even for high SNR. The proposed algorithm attains the CRLB for lower SNR than any of the algorithms used for comparison. 
   
   In the second scenario, we fixed the signal-to-noise ratio to $SNR = 10$ dB and evaluated the performance of algorithms for the different number of snapshots. From Fig.\ \ref{fig:tdu:snap}, it is seen that the RMSE of TD estimation decreases with the number of snapshots. It can be observed that the number of snapshots needs to be sufficiently high, i.e., equal or higher than $10$, for the algorithms to perform well, which is the consequence of the errors introduced by signal subspace estimation. 
   
   The same simulation scenarios are repeated for the case when the noise power is nonuniform across the receiver branches, and the corresponding RMSEs are shown in Fig.\ \ref{fig:tdnu:snr} and Fig.\ \ref{fig:tdnu:snap}. The signal-to-noise ratios in the third and fourth band are set to $-3$ dB and $-4.7$ dB compared to the SNR of the x-axis on the Fig.\ref{fig:tdnu:snr}. Likewise, in the fourth scenario, the signal-to-noise ratios for the third and fourth bands are set to $12$ dB and $11.3$ dB, respectively. Due to the appropriate weighting of the cost function (\ref{eq:sub_fit_prob}), the proposed algorithm is still close to the CRLB also in the case of nonuniform noise.
    
    \section{Conclusions}
    In this paper, we proposed an algorithm for time-delay estimation from multiband channel measurements. Considering the channel impulse response as a sparse signal in the time domain, we have formulated time-delay estimation as a problem of parametric spectral inference from observed multiband measurements. The acquired measurements exhibit multiple shift-invariance structures, and we estimate time-delays by solving the subspace fitting problem. The solution to the problem is found efficiently using the variable projection method. Future directions aim towards evaluating the proposed algorithm with real channel measurements and solving the problem of joint time-delay estimation and calibration of RF chains.
    



\bibliographystyle{IEEEtran}
\bibliography{paper}

\end{document}